\documentclass[copyright,creativecommons]{eptcs}
\providecommand{\event}{FOCLASA 2010} 
\usepackage{breakurl}             

\usepackage[american]{babel}
\usepackage{array}
\usepackage{xspace}
\usepackage{times}
\usepackage{mathptmx}
\usepackage[dvips,final,pdftex]{graphicx}
\usepackage{url}
\usepackage{wrapfig}

\newcommand{\ruler}[1]{\rule[0.5ex]{#1}{0.5pt}}

\newcommand{\M}[1][] %
  {\mbox{\textsf{M}\raisebox{-.4ex}{{\hspace{.02em}\small\textsf{#1}}}}\xspace}
\newcommand{\N}[1][] %
  {\mbox{\textsf{N}\raisebox{-.4ex}{{\hspace{.02em}\small\textsf{#1}}}}\xspace}

\begin{document}

\title{Simplified Distributed Programming with Micro Objects}

\author{Jan-Mark S. Wams, Maarten van Steen
\institute{VU University Amsterdam}
}

\def\authorrunning{J.M.S.\ Wams \& M. van Steen}

\maketitle

\thispagestyle{plain}
\pagestyle{plain}

\begin{abstract}

Developing large-scale distributed applications can be a daunting task. Object-based environments have
attempted to alleviate problems by providing distributed objects that look like local objects. We advocate
that this approach has actually only made matters worse, as the developer needs to be aware of many intricate
internal details in order to adequately handle partial failures. The result is an increase of application
complexity. We present an alternative in which distribution transparency is lessened in favor of clearer
semantics. In particular, we argue that a developer should always be offered the unambiguous semantics of
local objects, and that distribution comes from copying those objects to where they are needed. We claim that
it is often sufficient to provide only small, immutable objects, along with facilities to group objects into
clusters.

\end{abstract}

\section{Introduction}

Developing large-scale distributed applications has demonstrated to be a difficult task. Although one may
argue that considerable progress has been made in supporting application developers, the fact alone that data,
processes, and control are distributed across a potentially very large network of computers introduces unique
problems that cannot be concealed. Many problems are related to the dependability of a distributed
application, including availability and reliability of components, but also to integrity and security of the
application as a whole~\cite{birman.k2005}. Furthermore, most distributed applications have been designed with
a high degree of concurrency in mind, which in turn can easily lead to intricate communication and
coordination patterns~\cite{hohpe.g2004,schmidt.d2000}.

Underlying many, if not all approaches toward simplifying distributed application development, is the idea that
we should hide the intricacies of distribution from application developers. In other words, we should make
distribution as transparent as possible. This thought has led to a myriad of paradigms with distributed
object-based programming as perhaps the most prevalent. More recently, we are witnessing modern variants of
this paradigm in the form of Web services~\cite{alonso.g2004} and service-oriented computing in
general~\cite{huhns.m2005}.

Common to all these approaches is the client-server computing paradigm, in which a client process requests a
server process to execute an advertized service on its behalf, and to return the result. Note that the
peer-to-peer computing paradigm is often just a variant of the client-server model: in that case peers are
client and server at the same time. However, because the successful remote execution of instructions can never
be guaranteed, the client will always need to be prepared to handle partial failures that characterize
distributed systems. Of course, this result is well-known~\cite{fischer.m85}, but despite the attention that
it has been given in the past in the context of practical programming and systems
development~\cite{guerraoui.r1999,lea.d1997,vogels.w1998,waldo.j97} its ramifications have so far still been
largely ignored, leaving the programmer to solve the problem when it occurs. Yet, again, recent findings on
the impossibility of combining consistency and availability in partitionable networks confirm that we have a
serious problem to address~\cite{gilbert.s2002}.

Given these inherent difficulties, we argue, as others have done before us, that we should no longer try to
hide what cannot be hidden. We claim that the whole idea of remote execution of instructions in the presence
of failures often increases the complexity of applications instead of making them simpler. Distributed
applications are not, and never will be the same as their nondistributed variants. What is needed are models
in which distribution is \emph{apparent} (and not \emph{transparent}), and with clear and well-understood
semantics. In this paper, we propose to radically abandon the remote-instruction model in favor of a model
that allows only for the execution of local instructions, and which minimizes synchronization between
dislocated processes.

Our approach has a number of far-reaching consequences. First and foremost, being able to executing only local
operations and not delegate instructions to remote servers implies a \emph{copy-before-use} model: if a
process wants to operate on a data object, that object will have to be fetched from somewhere. This also
implies that we need an efficient object-location mechanism. Minimizing synchronization between dislocated
processes can be best supported by avoiding the need to \emph{move} a fetched object; instead, it should
merely need to be \emph{copied}, which leads to potentially massive replication of data objects. To come to
scalable solutions, we should then prevent synchronization of replicas in the presence of updates, which can
be achieved by making objects \emph{immutable}.

In this paper, we present a simple, yet powerful programming model and system for developing distributed
applications. Our model is founded on local-only operations on immutable micro objects that can (and generally
will) be massively replicated across a distributed system. We demonstrate how this simple model can be used to
develop complex distributed data structures such as complete file systems and messaging applications. In
doing so, we do not claim that we are presenting the best solution to the problems that are tampering
distributed programming. Instead, we wish to fuel the discussion on distribution transparency, as we do
believe that it deserves much more attention than researchers and practitioners are currently giving it.

\section{The Micro Object}
\label{sec:mo}

\begin{wrapfigure}{l}{0pt}
  $$\includegraphics[scale=0.35]{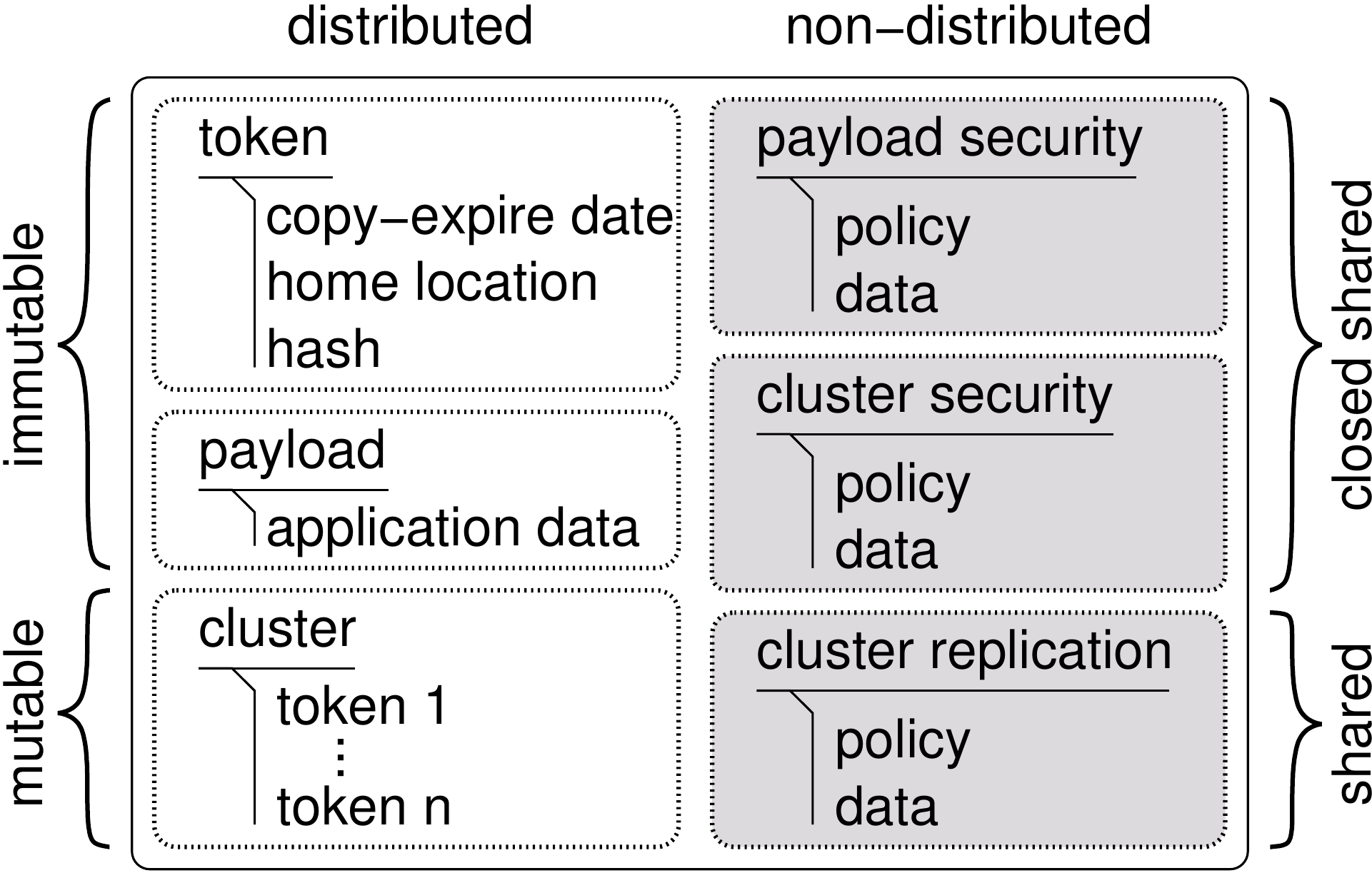}$$
  \caption{\label{fig:MoDi}The general organization of a micro object.}
\end{wrapfigure}

At the heart of our approach lies the notion of a \textbf{micro object}.  A micro object is a relatively small
container used to ferry copies of distributed data around. It should not be confused with traditional objects
from object-based programming in the sense that it is \emph{not} an encapsulation unit for data and associated
operations. For the creation of larger distributed data structures, micro objects can be clustered into
  arbitrary graphs.  The organization of a micro object is shown in Fig.~\ref{fig:MoDi}.

A micro object is used to distribute an immutable and a mutable data part. The immutable part comprises a
\textbf{token} by which the object can be uniquely identified and looked up, but more profound is the fact
that this section also consists of a limited-sized buffer of (encrypted) application data, or
\textbf{payload}. We stress that the payload cannot be modified, a design choice to which we return below.  To
accommodate at least some changes, a micro object can group related micro objects into a
\textbf{cluster}. Clusters allow for the construction of data structures as graphs, which we believe to be
strong enough for a huge class of applications.  Technically, a cluster is akin to an append-only list:
members can only be added, but never removed. As we discuss below, this restriction simplifies distribution
and the development of distributed applications.  These mutability constraints might seem too limiting,
however in Section~\ref{sec:cluster} we will show this not to be the case.  The immutable and mutable section
together form the distributed part of a micro object.  This part is copied and updated across a network by the
micro-object system, also referred to as the \textbf{MO system}.  An important issue is that the distributed
part is securely protected against unauthorized access.

Equally important is that the application using a (copy of a) micro object stays in full control regarding the
replication of the mutable part (i.e., the cluster) of that micro object.  An application (programmer) only
has to express the policy for replicating the object's cluster according to its own local demands.  For
example, if rapid dissemination is needed, an application may specify that a cluster should be flooded
throughout the network.  Whether flooding actually takes place depends on the (again local) needs of potential
recipients. We return to these issues below.

This protection and control is achieved through the nondistributed part of a micro object.  The nondistributed
part consists of two sections. The closed shared section describes how the payload and cluster sections of the
micro object are protected. Typically, this section contains policy descriptors and encryption keys;
information that may only be disclosed within a closed group through secure channels.  The openly shared
section describes the rules (i.e., the replication policy) that should be followed when copying (changes in)
cluster information to and from other address spaces.  By its nature, replication data has to be shareable,
however, it does not classify as distributed data, because it does not have to be the same for every
individual copy of a given micro object.  As we discuss below, these local policies provide a high degree of
flexibility in distributing and replicating micro objects.

\section{Example Scenario}
\label{sec:scenario}

To illustrate the organization and usage of micro objects, consider the following simple scenario.  Alice,
Bob, and Clare regularly publish news items that they would like to share (as micro objects) over a longer
period of time.  To this end, Alice takes the initiative to create a long-lasting micro object \M for storing
their shared news items.  In doing so, Alice's local \textbf{MO server} becomes the \textbf{home location} for
\M, effectively allowing others to be able to retrieve a copy of \M from that server.  To keep matters simple,
the home server's contact address is encoded in \M's token. We denote Alice's local copy of \M as \M[A].
Alice then passes the token of \M to Bob and Clare, using any out-of-band communication method, and
Bob and Clare retrieve their copy of \M (referred to as \M[B] and \M[C], respectively) from the MO
system. If we consider the who-knows-who graph based on the information contained in \M's token, we obtain the
situation as sketched in Fig.~\ref{fig:ReCa}(a): the MO server of Bob and Clare, respectively, know only the
MO server of Alice.

\newcolumntype{I}{!{\vrule width 1.5pt}}
\newlength{\savedwidth}
\newcommand{\whline}{\noalign{\global\savedwidth\arrayrulewidth\global\arrayrulewidth 1.5pt}\hline\noalign{\global\arrayrulewidth\savedwidth}}

\newcommand{\PBS}[1]{\let\temp=\\#1\let\\=\temp}
\newcommand{\RRCOL}{\PBS\raggedright\hspace{0pt}}
\begin{figure}
  \begin{center}
    \begin{tabular}{|c|>{\RRCOL}p{0.3\textwidth}|>{\RRCOL}p{0.3\textwidth}|}\hline
      & \textbf{Action} & \textbf{Effect} \\ \whline
      (a)
      \begin{minipage}{0.25\textwidth}
        \includegraphics{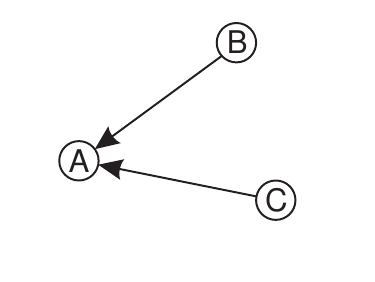}
      \end{minipage}
      &
      Alice passes \M to Bob and Clare &
      Bob and Clare know about the MO server of Alice \\ \hline
      (b)
      \begin{minipage}{0.25\textwidth}
        \includegraphics{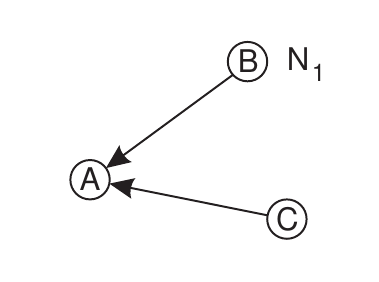}
      \end{minipage}
      &
      Bob creates micro object \N[1] and adds its token to \M[B] &
      Token of \N[1] is contained in \M[B] \\ \hline
      (c)
      \begin{minipage}{0.25\textwidth}
        \includegraphics{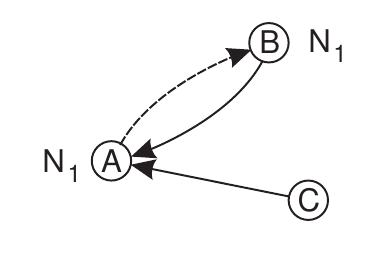}
      \end{minipage}
      &
      Bob's MO server attempts to flood elements contained in \M[B]'s cluster &
      Token of \N[1] is passed on to Alice; Alice's MO server learns about Bob's MO server \\ \hline
      (d)
      \begin{minipage}{0.25\textwidth}
        \includegraphics{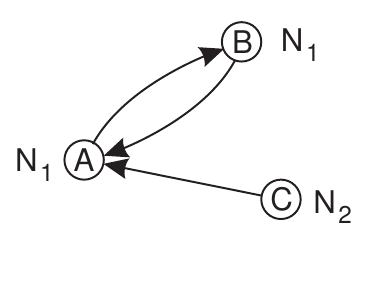}
      \end{minipage}
      &
      Clare creates micro object \N[2] and adds its token to \M[C] &
      Token of \N[2] is contained in \M[C] \\ \hline
      (e)
      \begin{minipage}{0.25\textwidth}
        \includegraphics{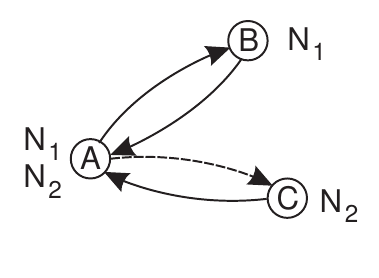}
      \end{minipage}
      &
      Clare's MO server attempts to flood elements contained in \M[C]'s cluster &
      Token of \N[2] is passed on to Alice; Alice's MO server learns about Clare's MO server \\ \hline
      (f)
      \begin{minipage}{0.25\textwidth}
        \includegraphics{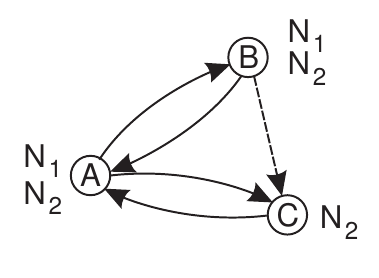}
      \end{minipage}
      &
      Alice's MO server attempts to flood elements contained in \M[A]'s cluster &
      Token of \N[2] is passed on to Bob; Bob's MO server learns about Clare's MO server \\ \hline
      (g)
      \begin{minipage}{0.25\textwidth}
        \includegraphics{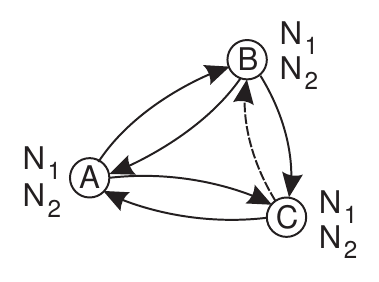}
      \end{minipage}
      &
      Bob's MO server attempts to flood elements contained in \M[B]'s cluster &
      Token of \N[2] is passed on to Clare; Clare's MO server learns about Bob's MO server \\ \hline
    \end{tabular}
    \caption{\label{fig:ReCa}The dissemination of a micro object over time. The replication policy has been
      set to \textsc{flooding}.}
  \end{center}
\end{figure}

The payload of \M will not need to hold any data other than perhaps a description of the type of news items it
is intended to contain.  In order to express that additions to the cluster of \M should be actively forwarded
to other parties, Alice, Bob, and Clare decide to set the replication policy of their local copies of \M to
\textsc{flooding}.  Now assume that Bob produces a news item that he wants to share with Alice and Clare.  To
that end, he creates a micro object (\N[1] at his local MO server) containing the actual news and adds the
object's token to his local copy \M[B] of \M, as shown in Fig.~\ref{fig:ReCa}(b).

At that point, the MO server storing \M[B] will make an attempt to forward any of the elements contained in
the object's cluster, as shown in Fig.~\ref{fig:ReCa}(c).  The only server it knows about, is the home
server of \M, i.e.\ Alice's local MO server.  Bob's server will then contact all the servers in the replication
data of \M[B], in this case only the home server of \M, to report the additions to the cluster of \M.  This
reporting is done by means of an \textsc{assent} request, which essentially initiates a harmonization of
cluster elements between \M[A] and \M[B]. From there on, the dissemination of the token proceeds as shown in
Fig.~\ref{fig:ReCa}, and similar for other newly created (tokens of) news items.

There are a number of important observations to make.  First, note that cluster elements are only tokens.  As
a consequence, after the clusters of \M[A], \M[B], and \M[C] have been merged (or, more strictly, harmonized),
the servers of Alice, Bob, and Clare will still need to explicitly fetch \N[1] or \N[2] to get (the payload
of) the new messages.  Also note that that news items can be forwarded only to MO servers that are known to
the forwarder \emph{and} that have indicated that they are willing to accept such items by means of a matching
replication policy.  An important effect of this need for matching is that, for example, Bob cannot produce a
news (or any other) item that will be stored at Alice's, Clare's or any other server without cooperation of
that server.

Finally, we point out that, in our example, the clusters of the local copies of \M did converge, as
distributed data should.  However, the replication data of the local copies of \M did not (need to) converge.
In effect, only after some elapse of time did all news items reach all interested parties.

\section{Design Issues}
\label{sec:design}

The main goal of the MO system is to make it easier for programmers to design and develop distributed
applications.  We claim that the MO system makes it easier to \emph{identify}, \emph{locate}, \emph{delete},
\emph{update}, \emph{protect}, and \emph{replicate} distributed data by providing a clear and singular way of
dealing with these issues.  Some of the protection and replication aspects, however, depend on local
(temporary) circumstances, and have to be dynamically directed by the application.  To this end, the MO system
offers a limited number of security and replication policies for the application (programmer) to choose from
that can be tuned by changing local security and replication data.  The replication data is shared throughout
the MO system when necessary, the security data, however, is \emph{closed shared data}, because it is shared
only in a closed group.  Furthermore, the MO system has no data access control, but is able to detect bogus
data to some extent.  We will elaborate on the design issues concerning all of these points below.

\subsection{Identifying and Locating Data}

Each micro object contains a systemwide unique token in order to simplify its processing in a highly
distributed environment. There are two important requirements for tokens. First, it should be relatively easy
to fetch a copy of an object given its token. Second, we need to ensure that a token always refers to the same
(unmodified) micro object.

Concerning fetching an object given only its token, in our current design each object has an associated
\textbf{home location} where it is \emph{guaranteed} to be available for copying until a specified
\textbf{copy-expire date}. The contact address of the home location as well as the copy-expire date are
encoded in an object's token, making an initial lookup extremely simple. Unless special measures have been
taken, more sophisticated lookup mechanisms will need to be used after the expire date, for example, as
deployed in peer-to-peer systems~\cite{risson.j2006}, or explicit location
services~\cite{steen1998.01}). Currently, we simply allow a home server to keep storing a micro object. Note,
however, that the original guarantees concerning the availability have actually expired.

A token also consists of a hash, which is computed over the home location, copy-expire date, the object's
payload, and a few other (smaller) fields.  Essentially, the hash ensures, with a high probability, that the
token is indeed systemwide unique, but is also uniquely associated with the payload, which, in turn, is
important for data integrity.  Note that a token can be computed locally; there is no need to communicate with
another party.

A consequence of this design is that the creator of a micro object is responsible for keeping it online until
its copy-expire date.  We do not consider this a drawback, but instead maintain it introduces a form of
fairness as data creators should now also provide the resources for keeping their data in the system.  In this
way, creators hold a bigger share in the cost of resources (CPU time, storage, network bandwidth) in
comparison to other approaches, like systems based on NNTP or SMTP.

Still, to make this home location scheme work, the system has to provide the means---until expiration---to
retrieve a copy of a given micro object from its token. Therefore, a home server needs to be always
online, just like the WWW depends on servers being online.  This scheme is simple, but not very robust.  To
compensate, the MO system contains additional replication options as we will describe below. As an
alternative, we have developed a system that allows for stable identifiers to be mapped to a possibly changing
collection of (home) servers~\cite{steen2007.10}. This alternative has not yet been integrated with the MO
system.

\subsection{Deleting Data}

Deleting a distributed object means deleting all its local copies. A delete operator could be fairly complex,
especially if it would need to guarantee that all replicas of an object had indeed been removed. To offload
the MO system from these issues so that we can keep it as simple as possible, we have decided to purposefully
not provide a delete operator. Instead, the only thing the MO system guarantees is that it will not remove an
object from its home location until its associated copy-expire date. In order to keep an object longer than
its copy-expire date, an application will need to explicitly take action, such as requiring its local server
to sustain the lifetime of the object. As we explain below, it can do so by specifying a local
\textsc{sustain} replication policy. A sustained object can still be located using the information in its
token.

To prevent premature copy expiration some form of clock synchronization between all participating parties is
needed.  The granularity of this synchronization need not be too fine and can easily be satisfied through a
time protocol such as NTP.  Assuming that the clock of a server can be kept up-to-date with a precision of
\,$T$ time units, a simple solution to premature expiration is to keep every micro object for a grace period
\,$T^* > T$ units after it's copy-expire date has ended.  Note that each server can locally determine its own
grace period based on the granularity and precision of its time synchronization mechanism.

To further simplify matters, an object may possibly also have a near-endless expiration date, effectively
implying that the MO system will never remove it from its home server. Such an approach is possible only if an
infinite lifetime of the home server can be guaranteed, or rather, that by using its address one can always
fetch a copy of the object. Such a scheme is not infeasible, as we have demonstrated when using mobile IPv6
addresses~\cite{steen2007.10}, yet it is well known that providing hard guarantees on the preservation of
objects is far from trivial~\cite{baker.m2006a}. We foresee that never deleting any data is a realistic,
viable option for systems such as ours, and that it may considerably contribute to keeping distributed
programming simple. However, in this paper we will not pursue this idea any further.

\subsection{Updating Data}

In all but the most trivial applications, data changes, and if the data is distributed, a local, cached or
replicated copy of that data might need to be updated.  One of the major challenges of any distributed system
is supporting timely propagation of updates of distributed data.  However, it is difficult, and often even
impossible for a system to predict which data will be updated, where updates will be needed, and when.  This
lack of knowledge is unfortunate, as better predictions will enhance the positive effects of replication, such
as responsiveness and availability.  Since even the application programmer often has a hard time predicting
changes, we separated the distributed part in a mutable and immutable part, as shown previously in
Fig.~\ref{fig:MoDi}.

This separation effectively concentrates changes in the mutable part of an object, making them better explicit
to both the application (programmer) as well as the MO system.  The mutable part (i.e., the cluster)
exclusively contains only tokens of micro objects.  Allowing only a set of tokens to change simplifies updates
considerably.  In our design, even the update operations on the mutable part are limited. In particular, there
is only an ``add-token'' operator and no ``remove-token,'' further simplifying the update process.

Moreover, the mutable part has been specially constructed for efficient replication by sorting its elements on
their copy-expire date.  This sorting allows us to construct efficient representations of clusters so that two
parties can quickly detect differences in their respective clusters. Note that since the copy-expire date is
part of the token, a list of tokens can be sorted locally, in line with our design philosophy.

This model forces the application (programmer) to express distributed application objects as immutable parts
glued together in a way that is efficient for distribution.  It can be argued that the combination of an
immutable payload and a limited mutable cluster is not enough to allow for distribution of arbitrary mutable
application data.  We advocate, however, that a broad range of fully mutable distributed application objects
can be efficiently supported.  In Section~\ref{sec:cluster}, we will substantiate this claim by means of an
example.

\subsection{Protecting Data}
\label{sub:PrDa}

The MO system supports fine granulated security of distributed data, because of the strict separation of
security management and object management. Note that different policies can be applied to securing an object's
payload and its cluster information. Distributing data raises fundamental security challenges.  The potential
number of people that could access distributed data could be huge and integrity and confidentiality of data
are not protected by personal hardware as is possible for nondistributed data.  Therefore, additional
protection is needed.  We opted for combined end-to-end encryption and authentication, because it
significantly lessens the security demands for remote parties.  Encryption prevents an attacker from reading
an object but does not protect against manipulating the data.  Authentication can protect against manipulation
of data but does not protect against reading of the data.

Note that although a combination of end-to-end encryption and authentication can be used to implement various
security policies, it does not always suffice.  Attacks based on traffic analysis could be repulsed by
sophisticated cryptographic protocols like mix-networks.

\subsection{Replicating Data}

As stated before, the MO system always utilizes a \emph{local copy} of a data object, where the traditional
approach is to utilize a \emph{remote copy} of a data object trough RPC or RMI.  This difference has important
implications for data replication.  In a traditional system, replication is deployed to enhance performance or
availability.  As a result, separate mechanisms are needed to support replica placement, consistency
enforcement, and redirecting clients to the best replica.  Moreover, replication may require the collaboration
of third-party servers, leading to the incentives and fairness problems hampering many of today's
decentralized peer-to-peer systems~\cite{vu.q2010}.

In a local-copy system such as the one we propose, purposefully replicating objects for availability and
performance can come at virtually no extra costs.  First, in order to access an object, an application will
have to make a local copy of that object.  We refer to this copying as basic replication. As a result, objects
are already replicated on demand to where they are needed. Combining basic replication with
\textsc{sustain}ing local copies and efficient lookup procedures beyond copy-expire dates, automatically
increases availability and access performance.

If an application strives for higher performance, robustness, or availability, it can specify this by means of
an additional replication policy, which is associated with the local copy of an object and its cluster. MO
servers with matching policies for the same object will then collaborate in (proactively) copying associated
clusters. An example of such a policy is \textsc{flooding}, which we discussed in
Section~\ref{sec:scenario}. Additional replication is established as an ad hoc agreement within a group of
collaborating local applications, whereas basic replication is supported by all MO servers, independent of
applications. In addition, as we explain later in Section~\ref{sec:cluster}, we allow for the specification of
a replication depth, i.e., to which level of referenced micro objects a replication policy should extend.

Having basic replication allows relaxation of demands put on the additional replication.  For example, assume
a group of applications jointly follow a gossip-based dissemination and replication of their objects by
applying an anti-entropy protocol~\cite{eugster.p2004}.  These protocols are known to disseminate data in a
robust way, but may easily introduce inconsistencies as different nodes will see a different set of objects.
Since the MO system can always rely on basic replication, these problems are alleviated when gossiping is used
as an additional way to replicate objects.  If the payload of a micro object is needed immediately, it can
always be fetched from its home server.

At first it might seem odd to allow a subset of MO servers to engage in an additional replication policy.  In
fact, we consider it one of the stronger points of the MO system that local copies of the same micro object
can have different replication policies.  For example, imagine a distributed file system based on the MO
system and assume that---at some point in time---a given file would be opened by a few of the participating
applications.  In this case, it would make perfect sense to let only those participating applications select a
high-cost, high-performance replication policy to keep the shared data structure (effectively consisting of
local clusters of replicated micro objects) consistent.

\section{Systems Design}
\label{sec:sys-design}

We will now discuss the design and parts of the implementation of our system.  The infrastructure of the MO
system is not unlike the \mbox{e-mail} system in that a distributed application does not directly contact
other applications.  Instead, a network of servers is used for distributing micro objects.

\begin{figure}
  $$\includegraphics[scale=0.35]{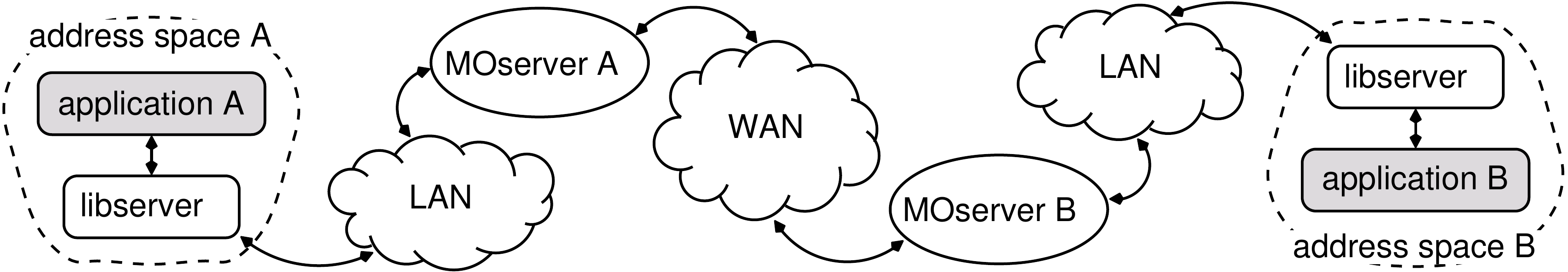}$$
  \vspace*{-24pt}
\caption{\label{fig:arch}Overall design of the MO system.}
\end{figure}

An application contacts a local server, much like an \mbox{e-mail} client application would do so for
transfering new messages between itself and a server provided by a company or ISP.  These servers will
communicate as peers to distribute micro objects.  Just like the \mbox{e-mail} system, an application can be
offline without disrupting any ongoing replication scheme.

Unlike the \mbox{e-mail} system, however, the MO system does not do end-point delivery.  In delivering
information, it is more like the WWW system: information is stored in a single known place and, possibly,
cached near the destination.  Like WWW proxy caching, multiple applications will generally be using the same
server cache for a better cache hit ratio.

On top of this basic ``pull on demand'' replication, the MO system features additional dynamic replication
policies.  The application (programmer) can specify when a server needs to spend additional resources on
replicating a specific micro object.

To handle basic and additional replication, the MO system follows the classical three-tier approach. The three
tiers in our implementation consists of the application, the lib-server, and the MO server (see
Fig.~\ref{fig:arch}).  The first tier, the application, shares its address space with the second tier, the
library server, also referred to as the lib-server.  The second tier, the lib-server, provides library
functions and spawns process threads acting like a server, hence the name.  The lib-server communicates with
the third tier, its local MO server, through a relative secure and fast connection, for example a LAN.  The MO
server has to be always online whereas the lib-server can be regularly offline.  In what follows we will take
a closer look at the MO server and the lib-server.

\subsection{The MO server}

\begin{wrapfigure}{l}{0pt}
  $$\includegraphics[scale=0.35]{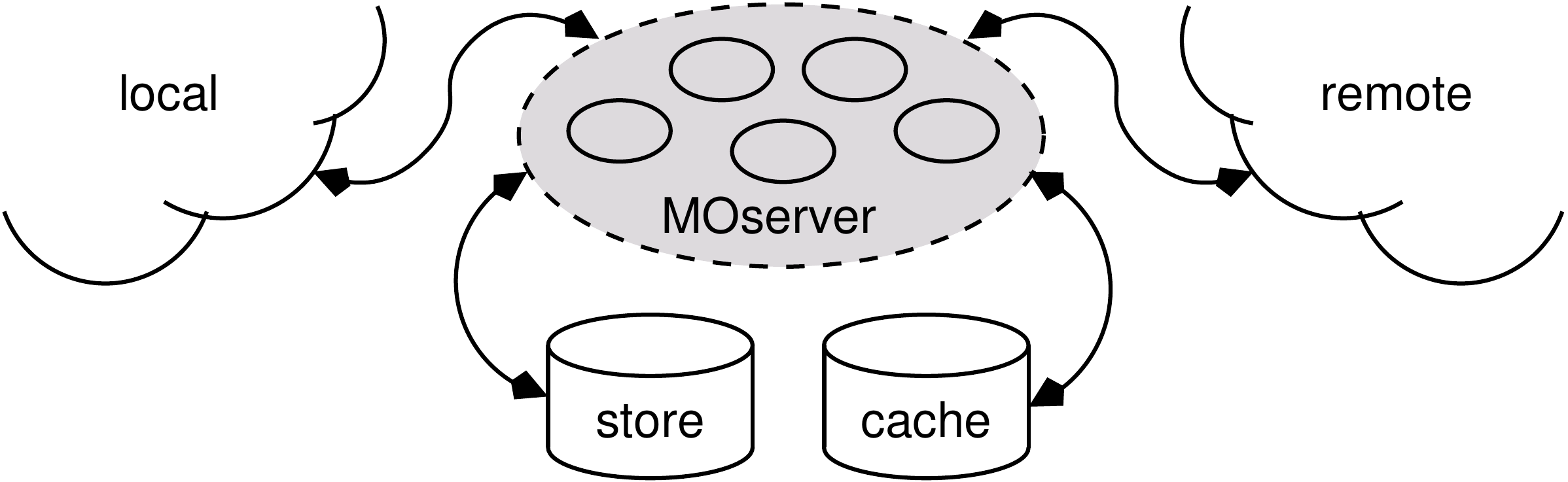}$$
  \vspace*{-24pt}
  \caption{\label{fig:serv}Close-up of the MO server.}
\end{wrapfigure}

The MO server, sketched in Fig.~\ref{fig:serv}, fulfills three major roles.  First, the MO server has to
store every micro object that a trusted MO-application has created.  The server will store such an object
until its copy-expire date, thus acting as the object's home server.  Second, it has to cache incoming micro
objects.  Third, it has to run threads to execute replication policies.

The store and cache differ mainly only in how they clean up their contents.  Micro objects can be removed from
the store only after their copy-expire date, while they can be removed from the cache at any time.  Just as with
additional replication, cache management has no external dependencies as a discarded micro object can always
be retrieved from its home server.  An MO server has a remote and a local communication channel. The local
channel differs from the remote channel in the sense that we assume it can be made as trustworthy as needed,
for example, by means of strong encryption.  In practice though, the local channel will simply be a LAN or ISP network
offering low latency and possibly also high bandwidth.  The MO server is a basic request/response system.  We
will refer to a request through the local channel as a local request, and to requests through the remote
channel as remote requests.

A remote request/response sequence is used by the MO servers to communicate with their remote peers.  There
are several types of such communications.  For example, if one MO server needs a micro object, it can ask any
other MO server for it by sending the latter a \textsc{fetch}-request containing a valid token.  If the receiving
MO server has the requested micro object (in its cache or store), it can send the micro object back in
response.  Since encryption is used at a higher level, there is no need for security checks, most notably
there is no distributed infrastructure for security.  It will also be difficult to forge a valid token, mostly
due to the sparsity of the token space.

To facilitate load balancing and additional replication policies a remote response can contain further
information by means of a \textbf{ditto-list}.  A ditto-list contains a number of MO servers that are likely to
give the proper response.  In general, an MO server on a ditto-list has previously made a similar request and
may therefore have relevant information to generate a proper response to the request.

Any remote request can trigger a \textsc{busy}-response with a ditto-list.  This reply indicates that the MO
server is swamped with similar requests.  It is then up to the requesting MO server to re-route the remote
request to another MO server.  Note that this solution is now sometimes applied to alleviate hot-spot problems
in the Web (see, e.g.,~\cite{patel.j2004}).

The \textsc{assent} request is sent whenever two MO servers want to make their respective copies of a micro
object consistent, i.e., make sure that the two associated clusters are harmonized.  To this end, an MO server
\textsf{A} can send an \textsc{assent}-request to MO server \textsf{B} containing micro object \M.  This
request will allow \textsf{B} to possibly merge the elements contained in \textsf{A}'s copy of \M's cluster
with its own copy of \M's cluster.  \textsf{B} can now also detect which elements are missing from
\textsf{A}'s copy of \M's cluster and pass this information back to \textsf{A}.  If both \textsf{A} and
\textsf{B} decide to add the missing elements to their respective copies, the two will be the same after the
\textsc{assent} exchange.  After merging, \textsf{A} or \textsf{B} might decide to forward information to
other servers, as we have seen in the example in Section~\ref{sec:scenario}.  Note, however, that each party
is completely free to decide which elements to include in its local copy of \M. As clusters may be very large,
efficiency of handling \textsc{assent} requests is essential.

A local request/response sequence is used by the MO server to communicate with a (trusted) lib-server.  An
obvious local request is \textsc{request-payload}, which is issued by the lib-server.  It forces the MO server
to get the payload of a requested micro object, either in its cache or store or by means of a remote fetch
request.  If the MO server is the home for the requested micro object (and its copy-expire date has not been
exceeded), it will---by definition---find the micro object in its store.  If not, the MO server can use a
remote \textsc{fetch}-request to a peer MO server, most notably the home MO server of the requested micro
object.  It will forward the response to the lib-server, but also extract the micro object from the response
and put it in its cache.  Note that in this case the MO server acts like a proxy server.  As with the
\textsc{fetch}-request, there is no need to check for access permissions.

Only a trusted application can ask an MO server to adopt---become the home of---a micro object.  It does so by
sending a local \textsc{adopt}-request.  The MO server will---if local policies allow---put the micro object
in its store.

Also only a trusted application can ask an MO server to start (or stop) executing a replication policy for a
given micro object.  It does so by sending a local \textsc{replicate}-request to the server.  The MO server
will---again, local policies permitting---start the requested replication policy for the cluster of the given
micro object.  Note that several replication policies can be active at the same time for a given micro object.
Therefore, the MO server has to be able to handle replication data of multiple replication policies per micro
object.

Trusted applications are also allowed to send a local \textsc{update}-request.  Such a request contains one or
more tokens that are to be added to a given object's cluster.  If the micro object in question is in the cache
or store, its cluster is updated immediately.  Also if there are replication policies active for this micro
object, they are evaluated, because the arrival of new cluster members may necessitate some action.

The store of an MO server holds all the micro objects that are at home at that server.  However, the store can
be populated with ``foreign'' micro objects too.  To understand why, note that every replication thread has
full (i.e., both read and write) access to the store.  Consequently, a replication policy like \textsc{sustain},
by which an object is stored beyond its copy-expire date could put such a foreign micro object in the store.
The result would be that this foreign micro object will not be removed from the MO system until its extended
copy-expire date.  Future replication strategies might have other reasons to put micro objects in the store,
for example, to save them from cache cleanups.  Note that every MO server can have its own policies for
storage, most notably it could feature a quota system, disallowing or charging excessive usage.

Since every MO server is also a proxy server---in that each application requests all its micro objects through
a local MO server---all MO servers feature a micro-object cache.  Appropriate caching algorithms for MO
servers still need to be investigated in detail.  For now, we have adopted an LRU algorithm.  Note that the
caching algorithm is a local affair, every MO server can make its own local decisions.  For example, it could
decide to cache requested micro objects dependent on which application issued the request.

We already mentioned the replication policy \textsc{sustain}. This replication policy is special because it
postpones the expiration of a micro object past its copy-expire date.  Basically, an application (programmer)
can ask an MO server to sustain a local copy of a micro object for a limited time (but not forever).  We
stress that an application needs to sustain the micro object at regular intervals (albeit that those intervals
may last long).  If a micro object is sustained on its home MO server, it will still be available to all other
MO servers.  If, however, a micro object is sustained on a set of MO servers not including the home MO server,
servers outside that set will not be able to fetch it anymore.  A prime candidate for prolonged sustaining,
for example, would be the root of a distributed file system.  Note that this does not imply that an
application has to be always online, but only frequently enough to prolong an object's lifetime.


\subsection{The Lib-Server}

The lib-server is linked into the application's address space as a library.  It provides the API of the MO
system.  Besides a library with functions, however, it also runs separate threads (in the background) in the
application's address space, acting like a server.  By putting the lib-server in the same address space as the
application, it has the same trust level.  This makes it simpler for the lib-server to safely access security
information like passwords.

\begin{wrapfigure}{l}{0pt}
  \begin{minipage}{0.7\textwidth}
  \begin{center}\small
    \begin{tabular}{@{~~}l@{~~}ll@{\hspace*{0.5cm}}l@{~~}l}
      \multicolumn{2}{c}{\ruler{4cm}}   &&  \multicolumn{2}{c}{\ruler{5.5cm}} \\
      home location    & (\verb|hloc_|) &&& \verb|plod_alloc(plod_t*);| \\
      copy-expire date & (\verb|xpir_|) &&& \verb|plod_put(plod_t*, size, buf);| \\
      token            & (\verb|tken_|) &&& \verb|plod_get_size(int*, plod_t);| \\
      payload          & (\verb|plod_|) &&& \verb|plod_get_store(char**, plod_t);|\\
      cluster          & (\verb|cter_|) &&& \verb|plod_free(plod_t*);| \\
      payload security & (\verb|psec_|) &&  \multicolumn{2}{c}{\ruler{5.5cm}} \\
      cluster security & (\verb|csec_|) &&  \\
      replication      & (\verb|repl_|) &&& \\
      micro object     & (\verb|mo_|)   &&& \\
      \multicolumn{2}{c}{\ruler{4cm}}   &&& \\
    \end{tabular}\vspace*{-12pt}
    \caption{\label{fig:adt} The lib-server ADT list and the payload interface.  }
  \end{center}
  \end{minipage}
\end{wrapfigure}

The application programming interface of the MO system (as implemented by the lib-server) consists of a
collection of abstract data types (ADTs), each with their own prefix, offering only a few ubiquitous library
functions.  Fig.~\ref{fig:adt}(a) lists the ADTs (with their prefix).  All ADTs, but the one for micro
objects, are relatively simple as illustrated by the ADT for the payload (\verb|plod_|), given in
Fig.~\ref{fig:adt}(b).  We discuss the internal working of the lib-server by describing the implementation
of the micro object (\verb|mo_|) ADT, which is given (in part) in Fig.~\ref{fig:mo}.

\begin{figure}[ht]
  \begin{center}\small
    \begin{tabular}{ll}
      \multicolumn{2}{c}{\ruler{0.95\textwidth}}\\
      \verb|mo_create_new(mo_t*, xpir_t, plod_t, psec_t, csec_t);| &  \verb|mo_cter_add_mo(mo_t*, mo_t);| \\
      \verb|mo_create_copy(mo_t*, tken_t, psec_t, csec_t);|        &  \verb|mo_put_repl(mo_t*, repl_t);| \\
      \verb|mo_put_cter_clbk(mo_t*, cter_t*, clbk_t, void*);|      &  \verb|mo_get_cter(cter_t*, mo_t);| \\
      \verb|mo_cter_wait(tken_t*, cter_t*, mo_t);|                 &  \verb|mo_get_plod(plod_t*, mo_t);| \\
      \verb|mo_cter_try_uwait(tken_t*, cter_t*, mo_t, long);|      &  \verb|mo_get_tken(tken_t*, mo_t);| \\
      \multicolumn{2}{c}{\ruler{0.95\textwidth}}\\
    \end{tabular}
  \end{center}\vspace*{-12pt}
  \caption{\label{fig:mo}List of the major micro object API calls.}
\end{figure}

All the API functions are thread-safe. The function \verb|mo_put_cter_clbk()| instructs the interface to
execute a given callback function, every time a new token is clustered to a given micro object. If the
programmer so chooses an application can also block and wait for new additions by calling
\verb|mo_cter_wait()|.

Since the lib-server threads can add tokens preemptively, the application needs a way to express what tokens
it considers ``old'' so the library server can present only the ``new'' additions. This holds for both the
call-back and the busy-wait functions. To this end a tracker cluster argument has to be supplied. Calling
either function with an empty tracker results in call-back function execution (or return from the wait
function) for every token that is already in, or consecutively added, to the cluster of the given micro
object.  Calling either function with a copy of the current cluster will trigger a response only to newly
added tokens. Also a \verb|mo_cter_try_uwait()| function is provided. This function either returns a newly
clustered token or \verb|NULL| if no new token was added after waiting for at least a given number of micro
seconds.

If or when the local lib-server will find out about a remote site adding an object to the cluster, is
dependent on the willingness of other MO servers in the system to cooperate and the local replication strategy
of \verb|myMo|.  Such cooperation, however, is likely to happen if applications of the same class are running
simultaneously (on different machines).

The lib-server also has separate threads that constitute the server part of the lib-server.  There are
three main reasons to add this server part.  First, without separate threads, the MO server would have to
resort to rendezvous communication, which would hinder performance.  Second, the combination of callback
functions and threads will also allow full multi-threaded applications.  Third, some cluster security policies
will not allow replication at the MO server level, so it has to be handled in the (trusted) application
address space, i.e., by the lib-server. We will not go into any details here.

\section{The Micro Object Clusters}
\label{sec:cluster}

\begin{wrapfigure}{r}{0pt}
  $$\includegraphics[scale=0.35]{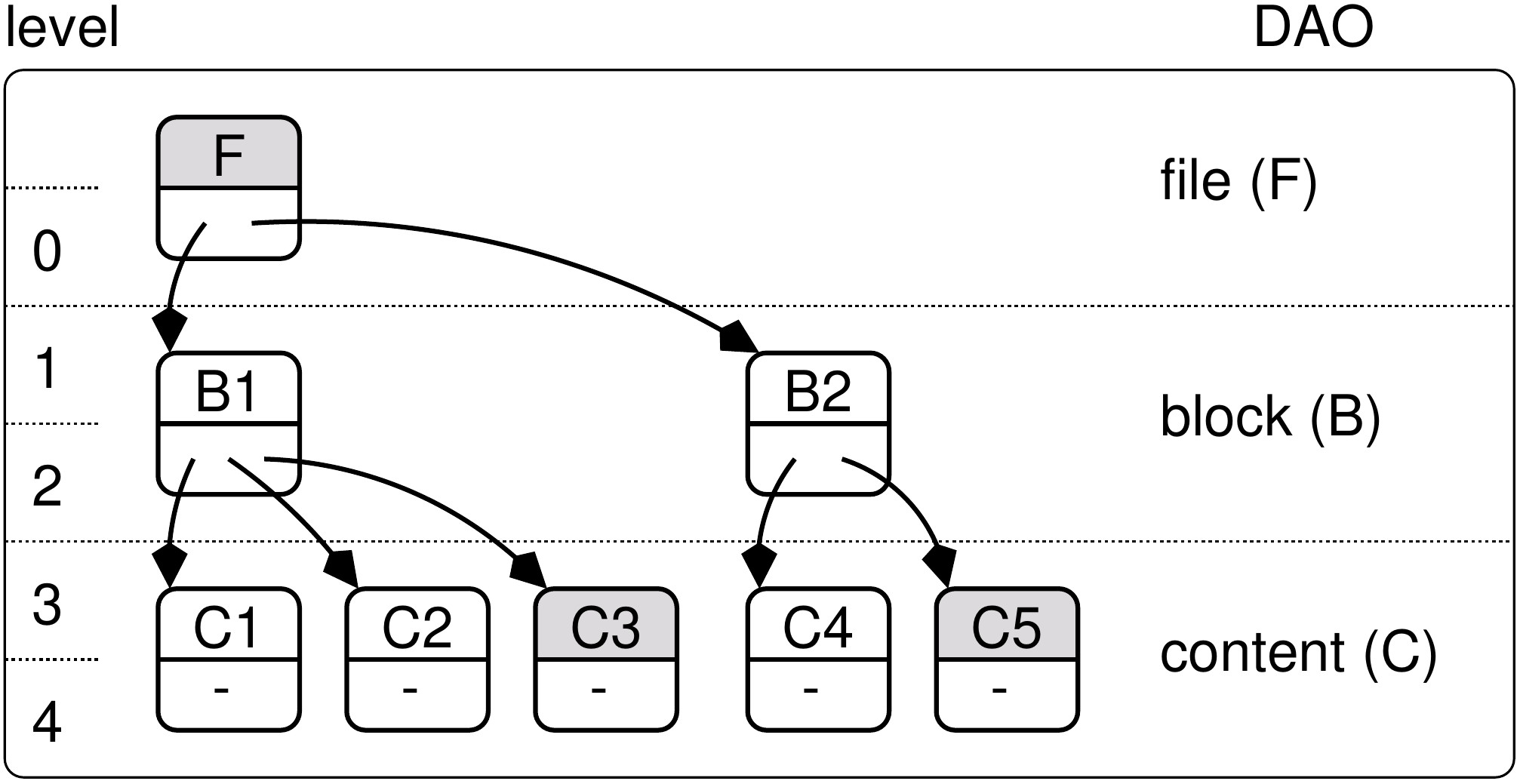}$$
  \caption{\label{fig:dfs}The realization of a distributed file of two blocks.}
\end{wrapfigure}

So far, we have shown that micro objects can be used to ferry application data and that clusters can be used
to build graphs of micro objects.  We will now demonstrate how micro object clusters can be used to construct
complex fully mutable \emph{distributed application objects} (DAOs).  In the MO system, an application
(programmer) defines every DAO as a {\em single} micro object with a (application specific) graph structure.
Sharing only a single micro object will nevertheless enable distributed applications to share a multitude of
objects that can be organized in any kind of graph.  Since every DAO is a micro object, complex DAOs can be
crafted by creating a micro object and adding one or more DAOs to its cluster.

As an example, consider the realization of a file DAO shared by a number of distributed file system
applications, shown in Fig.~\ref{fig:dfs}.  The cluster of the file DAO, \textsf{F}, contains two (tokens
of) block DAOs.  The cluster of the first block DAO, \textsf{B1}, contains three content DAOs.  The second
block DAO, \textsf{B2}, holds two content DAOs.  The content of a file DAO is defined as the concatenation of
the content of its clustered block DAOs, ordered by expiration date (\textsf{B1}, \textsf{B2} in this case).
The content of a block DAO is defined as (the payload of) the last content DAO from its cluster, ordered by
copy-expire date.  Thus the content of file \textsf{F} is the payload of \textsf{C3} followed by the
payload of \textsf{C5}.

From the MO system point of view, every DAO is a regular micro object and the structure of the graph
originating from its cluster has no meaning to the MO system.  One of the unique features of the MO system, is
that it still utilizes these graphs for grouping micro objects.  Grouping can be used to improve the
effectiveness of replication, especially if the objects are small.  This phenomenon is also known from other
fields, such as data clustering for efficient replication and distribution of databases~\cite{oszu.t99}.  By
using a replication level indicator in combination with a DAO, an application can generically inform the MO
system that a replication policy should be applied to all micro objects in a subgraph originating from a given
cluster.  The default value of the replication level is 0, to indicate that only the cluster itself should be
replicated.

To continue our example, assume that the sharing applications have set the replication policy of their copy of
\textsf{F} to \textsc{flooding} at level 3.  Consider what happens after one of the applications changes (the
second block of) its local copy of the file.  To update the file, a new block, \textsf{C6}, is constructed to
replace \textsf{C5}.  Next \textsf{C6} is added to (the cluster of) the local copy of \textsf{B2}.  Due to the flooding
level of \textsf{F}, \textsf{B2} (level 1 and 2) and the payload of \textsf{C6} (level 3) will be flooded, too.  Had
the level been set to 2, only the change would have been flooded (i.e., the cluster of \textsf{B2}), but not the
payload of \textsf{C6}.  Obviously, setting the level to 4 or higher, would not have made a difference.  We
stress that {\em the file DAO is fully mutable}, even though micro objects, themselves, are not.

The replication level is thus seen to provide the application (programmer) a simple yet powerful means to
express replication of larger groups of micro objects.

\section{Discussion}
\label{sec:discussion}

In this paper we have introduced a very different approach to distributed computing.  Instead of sending
messages to (possibly replicated) remote objects, we propose to let operations always take place on local
copies, keep data in objects immutable, and support only local graph-like data structures from which objects
can never be removed.  In our discussion so far, there are several ramifications of our approach that have
been barely touched upon.  Here, we briefly discuss two important ones: security and emergence.

\subsection{Security}

Building a secure large-scale distributed systems requires that security infrastructure is integrated into the
design from the start.  Therefore, the security infrastructure is natively incorporated into the MO system, as
illustrated in Fig.~\ref{fig:MoDi}.  The MO system needs data security and system security.  \emph{Data
security} is there to protect micro objects from unauthorized access, but also to protect applications against
bogus micro objects.  \emph{System security} concentrates on serving benign applications, while denying service
to malicious applications.

For data security, the MO system provides separated security policies that utilize (but are not limited to)
end-to-end encryption and authentication.  All sensitive security data is confined to the application address
space.

Bogus micro objects can be detected by end-to-end authentication.  However, a bogus micro object will be
detected only at the highest (i.e., application) level.  Therefore, bogus micro objects still threaten the
functionality of the lower level (i.e., the MO system itself).  To deal with DoS attacks, the MO system has
been designed such that most bogus data can be detected early.  Note that it is quite easy to generate a bogus
micro object and then calculate the correct hash value for its token.  However, it is unlikely that some
application would ever request such a micro object.  Generating a bogus micro object in response to a specific
request is computationally much harder, because the hash of the requested micro object is given as part of the
request.

The system security of the MO system is still subject to further research.  The MO system has rudimentary
protection against abuse of storage and transport.  In principle, MO servers can be tricked in to storing
bogus data, but it will end up in the cache so that the harm is limited.  A set of MO servers can sometimes be
tricked into transporting bogus data, however, newly developed policies and security for replication might
remedy this.  The MO system does not yet have protection (other than its hot spot handling), against denial of
service.  For example, a flooding attack will put parts of the MO system out of function.  Also, the MO system
suffers from the security bootstrapping problem: in order to set up secure communication between two given
parties, some pre-existing shared secret is needed.  Flooding and bootstrapping are common security problems,
and they are not specific to the MO system nor is it clear that these problems can be solved by changing the
design.

As mentioned in Section~\ref{sub:PrDa}, the MO system does not (yet) posses
any data flow shielding, and may thus leak sensitive data.

\subsection{Emergent Behavior}

Our emphasis on local decision making has important ramifications for overall system behavior.  For example,
as we explained, objects can be replicated across the system only if local policies of initiating and intended
peers match.  In contrast, replication in virtually all traditional distributed systems is based on explicit
and centralized control.  The effect of having only local policies is that we will see much more emergent
behavior, observed as the flow of (copies of) micro objects between servers.

It remains to be seen to what extent this emergent behavior can actually be controlled.  One avenue that we
are currently exploring is developing various replication policies and to see how combinations affect the
replication and distribution of micro objects.  Although the loss of centralized control can be seen as a
disadvantage, we believe that local decision making simplifies development and will certainly lead to much
better scalable solutions.

In this light, our approach is to be compared to the recent increase in gossip-based solutions, which all
evolve around local decision making~\cite{eugster.p2004}.  These solutions have in common that only by fine
tuning local decision rules can one observe desirable global behavior.  Unfortunately, the relation between
this local tuning and global behavior is often not well understood, and only recently have studies been
published in which different approaches are systematically compared~\cite{steen2007.15}.  However, it is also
clear that local decision making has excellent scalability properties, allowing systems to easily grow to
millions of nodes.  This point has already been demonstrated by traditional decentralized systems such as
those for exchanging news and e-mail.

\section{Conclusions}
\label{sec:conclusions}

Current message-to-object based distributed frameworks are ignoring partial failures.  As an alternative, we
propose a simple and clean model for distributed computing, which evolves around local decision making.  Our
design and prototype implementation indicate that we are dealing with a simple-to-realize model.  However, it
is yet too soon to draw hard conclusions on the viability of our proposal, although it is clear that it
contains the essential elements to tackle the hard problems that have been hampering large-scale distributed
systems.  Some of these hard problems, notably handling partial failures, are strongly alleviated by our
choice for combining local computing and immutability.  The drawback is some loss in distribution
transparency, a loss we believe is worthwhile taking.

It is clear we are only at the beginning of exploring this new paradigm.  For the immediate future, we will
concentrate our research efforts on, lib-server based cluster replication for enhanced security, ditto-list
population algorithms, and finding how many and which replication policies are practically needed.



\end{document}